\documentclass[conference]{IEEEtran}
\IEEEoverridecommandlockouts
\usepackage{cite}
\usepackage{amsmath,amssymb,amsfonts}
\usepackage{algorithmic}
\usepackage{textcomp}
\def\BibTeX{{\rm B\kern-.05em{\sc i\kern-.025em b}\kern-.08em
    T\kern-.1667em\lower.7ex\hbox{E}\kern-.125emX}}

\usepackage[dvips]{graphicx}
\usepackage{cite,amssymb,amsmath,epsfig,color}
\newcommand{\beq}{\begin{equation}}
\newcommand{\eeq}{\end{equation}}
\newcommand{\beqn}{\begin{eqnarray}}
\newcommand{\eeqn}{\end{eqnarray}}
\DeclareMathOperator*{\argmin}{arg\,min}
\def\bmath#1{\mbox{\boldmath$#1$}}

\begin{document}

\title{Performance Analysis of Distributed Radio Interferometric Calibration\\
\thanks{This work is supported by Netherlands eScience Center (project DIRAC, grant 27016G05).}
}

\author{\IEEEauthorblockN{Sarod Yatawatta\IEEEauthorrefmark{1}}
\IEEEauthorblockA{ASTRON,\\ The Netherlands Institute for Radio Astronomy,\\Dwingeloo, The Netherlands\\
Email: yatawatta@astron.nl\IEEEauthorrefmark{1}}
}

\maketitle

\begin{abstract}
Distributed calibration based on consensus optimization is a computationally efficient method to calibrate large radio interferometers such as LOFAR and SKA. Calibrating along multiple directions in the sky and removing the bright foreground signal is a crucial step in many science cases in radio interferometry. The residual data contain weak signals of huge scientific interest and of particular concern is the effect of incomplete sky models used in calibration on the residual. In order to study this, we consider the mapping between the input uncalibrated data and the output residual data. We derive an analytical relationship between the input and output probability density functions which can be used to study the performance of calibration.
\end{abstract}


\section{Introduction}
Most challenging science cases in modern radio astronomy are after weak signals  that are hidden under noise and bright foregrounds (see, e.g., \cite{SZ2013,Barry2016}). The main goal of calibration is the correction for systematic errors in the data and the removal of contaminating foregrounds from this data to reveal such weak signals. Consensus optimization \cite{boyd2011} has proved to be a computationally efficient solution for calibration \cite{DCAL,EUSIPCO2016,Brossard2018,CAMSAP2017,DMUX} as well as for imaging \cite{Onose,Meil2016,Degu2016,Onose2017} massive amounts of radio interferometric data. Calibration is always imperfect due to the errors in the input sky model as well as the consensus polynomials being used. Cramer-Rao lower bounds \cite{Zmu,Jeffs06,Wijn,Kazemi12} have been used to study the asymptotic variance of estimation error of calibration parameters. Translating this bound to the error in the residual is however, cumbersome. In order to overcome this, calibration is considered as a nonlinear regression and Jacobian leverage \cite{cook1982residuals,Laurent92,Laurent93} is proposed in \cite{SS9,Patil2016} to get limits on the variance of the residuals.

In this paper, we consider the mapping between the input uncalibrated data and the output residual data, where the residual is obtained after calibration and removal of the bright foreground signals. We derive an analytic relationship between the probability density functions (PDFs) of the input and output. In order to do that we use developments in bi-level optimization \cite{Samuel,Gould2016} and matrix differentiation \cite{Mdiff} to find derivatives of $\mathrm{argmin}$ function used in calibration. 

The rest of the paper is organized is as follows. In section \ref{sec:model}, we give an overview of distributed calibration in radio interferometry. In section \ref{sec:perf}, we derive analytic relationships for the performance of distributed calibration. Next, in section \ref{sec:pdf} we derive a relationship between the input and output PDFs and finally, we draw our conclusions in section \ref{sec:conc}.

{\em Notation}: Lower case bold letters refer to column vectors (e.g. ${\bmath y}$). Upper case bold letters refer to matrices (e.g. ${\bf { C}}$). Unless otherwise stated, all parameters are complex numbers. The matrix inverse, transpose, Hermitian transpose, and conjugation are referred to as $(.)^{-1}$, $(.)^{T}$, $(.)^{H}$, $(.)^{\star}$, respectively. The matrix Kronecker product is given by $\otimes$. The vectorized representation of a matrix is given by $\mathrm{vec}(.)$. The identity matrix of size $N\times N$ is given by ${\bf {I}}_N$.  Estimated parameters are denoted by a hat, $\widehat{(.)}$. All logarithms are to the base $e$, unless stated otherwise. The Frobenius norm is given by $\|.\|$.

\section{Radio Interferometric Calibration}\label{sec:model}
In this section we give a brief overview of the data model used in radio interferometric calibration \cite{HBS,TMS}. We consider the radio frequency sky that is part of the sky model to be composed of discrete sources, far away from the earth such that the approaching radiation from each one of them appears to be plane waves.  There are $N$ receiving elements with dual polarized feeds in the array and  at the $p$-th station, this plane wave causes an induced voltage, which is dependent on the beam attenuation as well as the radio frequency receiver chain attenuation. Consider the correlation of signals at the $p$-th receiver and the $q$-th receiver, at frequency $f$, with proper signal delay. After correlation, the correlated signal of the $p$-th station and the $q$-th station (named as the {\em visibilities}), ${\bf { V}}_{pqf}$ ($\in {\mathbb C}^{2\times 2}$) is given by 
\beq \label{vispq}
{\bf { V}}_{pqf}= {\bf { J}}_{pf} {\bf { C}}_{pqf} {\bf { J}}_{qf}^{H} + {\bf { N}}_{pqf}.
\eeq
In (\ref{vispq}), ${\bf { J}}_{pf}$ and ${\bf { J}}_{qf}$ ($\in {\mathbb C}^{2\times 2}$)  are the Jones matrices describing systematic errors at frequency $f$, at stations $p$ and $q$, respectively. These matrices represent the effects of the propagation medium, the beam shape and the receiver. The noise matrix is given as ${\bf { N}}_{pqf}$ ($\in {\mathbb C}^{2\times 2}$). The intrinsic signal on baseline $pq$ is given by the coherency matrix ${\bf { C}}_{pqf}$ ($\in {\mathbb C}^{2\times 2}$). For a linearly polarized source in the sky, with Stokes parameters $I_{pqf},Q_{pqf},U_{pqf},V_{pqf}$, we have 
\beq \label{coh}
{\bf { C}}_{pqf}=e^{\jmath \phi_{pqf}}\left[ \begin{array}{cc}
I_{pqf}+Q_{pqf} & U_{pqf}+\jmath V_{pqf}\\
U_{pqf}-\jmath V_{pqf} & I_{pqf}-Q_{pqf}
\end{array} \right]
\eeq
where $\phi_{pqf}$ is the Fourier phase component that depends on the direction in the sky as well as the separation of stations $p$ and $q$. For baseline coordinates $u_{pq},v_{pq},w_{pq}$ (in wavelengths) and direction cosines  $l,m,n$, we have $\phi_{pqf}=-2\pi(u_{pq}l+v_{pq}m+w_{pq}(n-1))$. The noise matrix ${\bf { N}}_{pqf}$ is assumed to have elements with zero mean, complex Gaussian entries with equal variance in real and imaginary parts but the statistics will vary because of the unmodelled structure \cite{Kaz3,SIRP,grobler2014}. 
The cost function that is minimized is given as
\beq \label{cost}
g_{f}({\bf J}_f)= \sum_{p,q}\| {\bf V}_{pqf} - {\bf A}_p{\bf J}_f {\bf C}_{pqf} ({\bf A}_q{\bf J}_f)^H \|^2
\eeq
where the systematic errors for all $N$ stations are grouped as ${\bf J}_f\in \mathbb{C}^{2N\times 2}$,
\beq
{\bf J}_f\buildrel\triangle\over=[{\bf J}_{1f}^T,{\bf J}_{2f}^T,\ldots,{\bf J}_{Nf}^T]^T.
\eeq
Using the canonical selection matrix ${\bf A}_p$ ($\in \mathbb{R}^{2\times 2N}$), where only the $p$-th block is ${\bf I}_2 \in \mathbb{R}^{2\times 2}$,
\beq \label{Ap}
{\bf A}_p \buildrel\triangle\over=[{\bf 0},{\bf 0},\ldots,{\bf I}_2,\ldots,{\bf 0}],
\eeq
we can select the systematic errors for the station $p$  as ${\bf A}_p{\bf J}_f$.
Note that in (\ref{cost}), the summation is taken over all the baselines $pq$ that have data, within a small bandwidth and time interval within which the systematic errors are assumed to be fixed.

Consensus optimization problem is formulated as follows. First, we create the augmented Lagrangian as
\beq \label{aug}
L_f({\bf J}_f,{\bf Z},{\bf Y}_f)=g_{f}({\bf J}_f) + \|{\bf Y}_f^H({\bf J}_f-{\bf B}_f {\bf Z})\| + \frac{\rho}{2} \|{\bf J}_f-{\bf B}_f {\bf Z}\|^2
\eeq
where the subscript $(.)_f$ denotes data (and parameters) at frequency $f$. In (\ref{aug}), $g_{f}({\bf J}_f)$ is the original cost function as in (\ref{cost}). The Lagrange multiplier is given by ${\bf Y}_f$ ($\in \mathbb{C}^{2N\times 2}$). The global variable ${\bf Z}$ ($\in \mathbb{C}^{2FN\times 2}$) is shared by data at all $P$ frequencies. Consensus polynomial basis (with $F$ terms) is represented by the matrix ${\bf B}_f = {\bf b}_f^T\otimes {\bf I}_{2N}$ ($\in \mathbb{R}^{2N\times 2FN}$) with ${\bf b}_f$ ($\in \mathbb{R}^{F\times 1}$) representing the basis functions evaluated at frequency $f$. The regularization parameter is given by $\rho$ ($\in \mathbb{R}^{+}$).
The alternating direction method of multipliers (ADMM) iterations $n=1,2,\ldots$ for solving (\ref{aug}) are given as
\beqn \label{step1}
({\bf J}_f)^{n+1}= \underset{{\bf J}}{\argmin}\ \ L_f({\bf J},({\bf Z})^n,({\bf Y}_f)^n)\\ \label{step2}
({\bf Z})^{n+1}= \underset{{\bf Z}}{\argmin}\ \ \sum_f L_f(({\bf J}_f)^{n+1},{\bf Z},({\bf Y}_f)^n)\\ \label{step3}
({\bf Y}_f)^{n+1}=({\bf Y}_f)^n + \rho\left( ({\bf J}_f)^{n+1}-{\bf B}_f ({\bf Z})^{n+1} \right)\\ \label{step4}
\eeqn
where we use the superscript $(.)^n$ to denote the $n$-th iteration where (\ref{step1}) to (\ref{step4}) are executed in order. The steps (\ref{step1}),(\ref{step3}) and (\ref{step4}) are done for each $f$ in parallel, at each compute (slave) node. The slave nodes are distributed across a network of computers. The update of the global variable in (\ref{step2}) is done in closed form at the fusion center. 
The extension of this data model to a multi-source scenario can be found in e.g., \cite{DMUX}.

\section{Performance analysis}\label{sec:perf}
At convergence, the closed form update of the global variable ${\bf Z}$ is
\beq \label{Zest}
{\bf { Z}}=\left( \sum_i \rho {\bf { B}}_{f_i}^T {\bf { B}}_{f_i} \right)^{\dagger} \left(\sum_i {\bf { B}}_{f_i}^T ({\bf { Y}}_{f_i}+\rho {\bf { J}}_{f_i})\right)
\eeq
where ${\bf { B}}_{f_i}$ corresponds to the consensus polynomial terms evaluated at frequency $f_i$. We separate one frequency $f_i=f$ from the other $P-1$ frequencies to get
\beq \label{Zest1}
{\bf Z}= \rho {\bf P}{\bf J}_f+{\bf P}{\bf Y}_f+{\bf R}
\eeq
where
\beq
{\bf P} \buildrel \triangle \over=\left( \sum_i \rho {\bf { B}}_{f_i}^T {\bf { B}}_{f_i} \right)^{\dagger} {\bf { B}}_{f}^T\ \ \in\mathbb{C}^{2FN\times 2N},
\eeq
and
\beq\label{remainder}
{\bf R} \buildrel \triangle \over= \left( \sum_i \rho {\bf { B}}_{f_i}^T {\bf { B}}_{f_i} \right)^{\dagger} \left(\sum_{i,f_i\ne f} {\bf { B}}_{f_i}^T ({\bf { Y}}_{f_i}+\rho {\bf { J}}_{f_i})\right). 
\eeq 
Note that ${\bf R}$ ($\in\mathbb{C}^{2FN\times 2}$) in (\ref{remainder}) has no dependence on the variables at frequency $f$, i.e., ${\bf J}_f$ and ${\bf Y}_f$. Substituting (\ref{Zest1}) to (\ref{aug}), we get
\beqn \label{Lconverge}
\lefteqn{L_f({\bf J}_f,{\bf Y}_f)}\\\nonumber
&&=g_{f}({\bf J}_f)+\|{\bf Y}_f^H\left(({\bf I}-\rho{\bf B}_f{\bf P}){\bf J}_f-{\bf B}_f{\bf P}{\bf Y}_f-{\bf B}_f{\bf R}\right)\|\\\nonumber
&&+\frac{\rho}{2}\|({\bf I}-\rho{\bf B}_f{\bf P}){\bf J}_f-{\bf B}_f{\bf P}{\bf Y}_f-{\bf B}_f{\bf R}\|^2.
\eeqn
The gradients of (\ref{Lconverge}) with respect to  ${\bf J}_f,{\bf Y}_f$ are given as
\beqn \label{gradJ}
\lefteqn{{\rm grad}(L_f,{\bf J})}\\\nonumber
&&={\rm grad}(g_{f}({\bf J}_f),{\bf J}_f) + {\bf F}^H{\bf F}\frac{\rho}{2}{\bf J}_f+{\bf F}^H{\bf F}\frac{1}{2}{\bf Y}_f+ {\bf r}_1({\bf R}_1)
\eeqn
and
\beq \label{gradY}
{\rm grad}(L_f,{\bf Y})=\frac{1}{2}{\bf F}^H{\bf F}{\bf J}_f-\frac{1}{2\rho}({\bf I}-{\bf F}^H{\bf F}){\bf Y}_f+{\bf r}_2({\bf R}_2)
\eeq
where ${\bf F}\buildrel \triangle \over={\bf I}_{2N}-\rho{\bf B}_f{\bf P}$  ($\in \mathbb{C}^{2N\times 2N}$) and ${\bf r}_1({\bf R}_1)$,${\bf r}_2({\bf R}_2)$ are the remainder terms that are independent of ${\bf J}_f$ and ${\bf Y}_f$. The proof is given in appendix I.
The gradient of the original cost function (\ref{cost}) is
\beqn
\lefteqn{ {\rm grad}(g_{f}({\bf J}_f),{\bf J}_f)}\\\nonumber
&=&-\sum_{p,q} \left({\bf { A}}_p^T({\bf { V}}_{pqf}-{\bf { A}}_p {\bf { J}}_f {\bf { C}}_{pqf} {\bf { J}}_f^H {\bf { A}}_q^T){\bf { A}}_q{\bf { J}}_f{\bf { C}}_{pqf}^H \right.\\\nonumber
&& + \left. {\bf { A}}_q^T({\bf { V}}_{pqf}-{\bf { A}}_p{\bf { J}}_f{\bf { C}}_{pqf}{\bf { J}}_f^H{\bf { A}}_q^T)^H {\bf { A}}_p {\bf { J}}_f{\bf { C}}_{pqf}\right)
\eeqn
and the derivation and be found in \cite{DCAL}.
At a local minimum, we have 
\beqn \label{lmin1}
{\rm grad}(L_f,{\bf J})={\bf 0},\\ \label{lmin2}
{\rm grad}(L_f,{\bf Y})={\bf 0}. 
\eeqn
Consider $x_{p^\prime q^\prime r} \in \mathbb{R}$ to be one data point out of many that completes a full observation ${\bf V}_{pq}$, $p,q\in[1,N]$,$p\ne q$. This data point belongs to $p=p^\prime,q=q^\prime$ correlation pair and $r\in[1,8]$. We select the value of $r$ to represent one real or imaginary value of  ${\bf V}_{p^\prime q^\prime}$ ($\in {\mathbb C}^{2\times 2}$). Note that each complex number is considered as two data points. For instance, if $r=1$, we represent the real part of ${\bf V}_{p^\prime q^\prime}(1,1)$. If $r=2$, the imaginary part of ${\bf V}_{p^\prime q^\prime}(1,1)$ is selected, and so on. 

In order to find $\frac{\partial {\bf J}}{\partial x_{p^\prime q^\prime r}}$, we take the derivative of both sides of (\ref{lmin1}) and (\ref{lmin2})  as in \cite{Samuel,Gould2016} and we get
\beqn \label{Jderiv}
\lefteqn{\mathrm{vec}\left(\frac{\partial {\bf J}_f}{\partial x_{p^\prime q^\prime r}}\right)}\\\nonumber
&&=\left( \mathcal{D}_{\bf J}{\rm grad}(g_{f}({\bf J}_f))\right. \\\nonumber
&&\left. + \frac{\rho}{2}{\bf I}\otimes\left({\bf F}^H{\bf F}\left({\bf I}+\left({\bf I}-{\bf F}^H{\bf F}\right)^{-1}{\bf F}^H{\bf F}\right)\right)\right)^{-1}\\\nonumber
&&\times \left({\bf A}_{q^\prime}{\bf J}_f {\bf C}_{p^\prime q^\prime f}^H\right)^T \otimes {\bf A}_{p^\prime}^T \mathrm{vec}\left(\frac{\partial {\bf V}_{p^\prime q^\prime}}{\partial x_{p^\prime q^\prime r}}\right)
\eeqn
where 
\beqn
\lefteqn{\mathcal{D}_{\bf J}{\rm grad}(g_{f}({\bf J}_f))=}\\\nonumber
&&\sum_{p,q} \left( -({\bf C}_{pqf}^{H})^{T}\otimes {\bf A}_p^T {\bf V}_{pqf}{\bf A}_q 
-{\bf C}_{pqf}^{T}\otimes {\bf A}_q^T {\bf V}_{pqf}^H{\bf A}_p  \right. \\\nonumber
&+&\left.({\bf C}_{pqf}{\bf J}_f^H{\bf A}_q^T{\bf A}_q{\bf J}_f{\bf C}_{pqf}^H)^T\otimes {\bf A}_p^T{\bf A}_p \right. \\\nonumber
&+&\left.({\bf C}_{pqf}^H{\bf J}_f^H{\bf A}_p^T{\bf A}_p{\bf J}_f{\bf C}_{pqf})^T\otimes {\bf A}_q^T{\bf A}_q \right. \\\nonumber
&+&\left. ({\bf C}_{pqf}^{H})^{T}\otimes {\bf A}_p^T{\bf A}_p {\bf J}_f{\bf C}_{pqf}{\bf J}_f^H{\bf A}_q^T{\bf A}_q \right. \\\nonumber
&+&\left. ({\bf C}_{pqf}^{T})\otimes {\bf A}_q^T{\bf A}_q {\bf J}_f{\bf C}_{pqf}^H{\bf J}_f^H{\bf A}_p^T{\bf A}_p \right).
\eeqn
The proof can be found in appendix II. Note that $\frac{\partial {\bf V}_{p^\prime q^\prime}}{\partial x_{p^\prime q^\prime r}}$ in (\ref{Jderiv}) will give a matrix ($\in {\mathbb C}^{2\times 2}$) with all zeros, except one real or imaginary value equal to $1$, depending on the value of $r$.

The residual is calculated by subtracting the calibrated model from the data as
\beq
{\bf R}_{pqf}={\bf V}_{pqf} - {\bf A}_p {\bf J}_f {\bf C}_{pqf} {\bf J}_f^H {\bf A}_q^T.
\eeq
We take the derivative of the residual with respect to $x_{p^\prime q^\prime r}$ and using \cite{Mdiff}, we get
\beqn \label{Rderiv}
 \mathrm{vec}\left( \frac{\partial{\bf R}_{pqf}} {\partial x_{p^\prime q^\prime r}} \right)=&& \mathrm{vec}\left( \frac{\partial{\bf V}_{pqf}} {\partial x_{p^\prime q^\prime r}} \right)\\\nonumber
&&-\left(\left({\bf C}_{pqf}{\bf J}_f^H {\bf A}_q^T\right)^T \otimes {\bf I}\right) {\bf A}_p \mathrm{vec}\left( \frac{\partial{\bf J}_{f}} {\partial x_{p^\prime q^\prime r} } \right).
\eeqn
Note that $\mathrm{vec}\left( \frac{\partial{\bf V}_{pqf}} {\partial x_{p^\prime q^\prime r}} \right)$ in (\ref{Rderiv}) is zero except when $p=p^\prime,q=q^\prime$. Using (\ref{Jderiv}) and (\ref{Rderiv}), we can study the behavior of the residual with respect to small changes in input data. In section \ref{sec:pdf}, we develop this further to consider the relationship between the input data and output residual PDFs.

\section{Probability Density Functions}\label{sec:pdf}
We reformulate (\ref{cost}) as a vector optimization problem, for the sake of simplicity. The vectorized form of (\ref{vispq}), ${\bf v}_{pq}=\mathrm{vec}({\bf { V}}_{pq})$  can be written as 
\beq \label{vecvispq}
{\bf v}_{pqf}= {\bf { J}}_{qf}^{\star}\otimes {\bf { J}}_{pf} \mathrm{vec}({\bf { C}}_{pqf}) + {\bf n}_{pqf}
\eeq
  where ${\bf n}_{pqf}=\mathrm{vec}(\bf { N}_{pqf})$. Depending on the time and frequency interval within which calibration solutions are obtained, we can stack up all cross correlations within that interval as
\beq
{\bf x}=[\mathrm{real}({\bf v}^T_{12f})\ \mathrm{imag}({\bf v}^T_{12f})\ \ldots \ldots \mathrm{imag}({\bf v}^T_{(N-1)Nf})]^T
\eeq
where  ${\bf x}$ is a vector of size $D\times 1$ of real data points. For a single time sample, $D=8N(N-1)/2$ because each (unique) cross correlation produces $8$ real data points. One element out of this vector is  $x_{p^\prime q^\prime r}$ (considered in section \ref{sec:perf}), where $p^\prime,q^\prime$ denote the pair of receivers forming the correlation and $r$ is one data point out of the $8$ produced by each correlation.   We have the data model 
\beq \label{obs}
{\bf x}={\bf s}({\bmath \theta}) + {\bf n}
\eeq
where ${\bmath \theta}$ is the real parameter vector (size $M\times 1$) that is estimated by calibration. The parameters ${\bmath \theta}$ are the elements of ${\bf { J}}_{pf}$-s, with real and imaginary parts considered separately.

The maximum likelihood (ML) estimate of ${\bmath \theta}$ under zero mean, white Gaussian noise is obtained by minimizing the least squares cost 
\beq \label{mltheta}
\widehat{\bmath \theta}=\underset{\bmath \theta}{\rm arg\ min} f({\bf x},{\bmath \theta}) 
\eeq
where 
\beq
f({\bf x},{\bmath \theta}) \buildrel\triangle \over =\|{\bf x}- {\bf s}({\bmath \theta})\|^2.
\eeq

The residual using calibration solution $\widehat{\bmath \theta}$ is obtained as
\beq \label{residual}
{\bf y}= {\bf x}-{\bf s}(\widehat{\bmath \theta}).
\eeq

 The CRLB \cite{Zmu,Jeffs06,Wijn,Kazemi12} is used to find a lower bound to the variance of $\widehat{\bmath \theta}$. However, relating this lower bound to the residual ${\bf y}$ is not simple. Using Jacobian leverage \cite{cook1982residuals,Laurent92,Laurent93}, it is possible to obtain limits for the variance of ${\bf y}$ \cite{SS9,Patil2016} but we are after a simpler approach.

For $m=[1,\ldots,D]$, consider $x_m$ to be one element of ${\bf x}$, and this is the same $x_{p^\prime q^\prime r}$ considered in section \ref{sec:perf}, except we use $m$ as the index instead of $p^\prime,q^\prime,r$. Taking the derivative of the residual with respect to $x_m$, we have
\beq \label{elem}
\frac{\partial {\bf y}} {\partial x_m}=\frac{\partial {\bf x}}{\partial x_m} - \frac{\partial}{\partial x_m} {\bf s}({\bmath \theta}) \arrowvert_{{\bmath \theta}=\widehat{\bmath \theta}}.
\eeq

Using the chain rule (at ${\bmath \theta}=\widehat{\bmath \theta}$)
\beq
 \frac{\partial}{\partial x_m} {\bf s}({\bmath \theta}) \arrowvert_{{\bmath \theta}=\widehat{\bmath \theta}} =\frac{\partial {\bf s}({\bmath \theta})}{\partial {\bmath \theta}^T} \arrowvert_{{\bmath \theta}=\widehat{\bmath \theta}} \times \frac{\partial \widehat{\bmath \theta}}{\partial x_m}
\eeq
where $\frac{\partial {\bf s}({\bmath \theta})}{\partial {\bmath \theta}^T}  \in \mathbb{R}^{D\times M}$ and $\frac{\partial {\bmath \theta}}{\partial x_m} \in \mathbb{R}^{M\times 1}$.

At the solution, the gradient of the cost function is zero, i.e.,  $\frac{\partial f({\bf x},{\bmath \theta})}{\partial {\bmath \theta}} \arrowvert_{{\bmath \theta}=\widehat{\bmath \theta}}  ={\bf 0}$. Following \cite{Gould2016,Samuel}, we have \footnote{Let $f^{\prime}({\bf x},{\bmath \theta})=\frac{\partial f({\bf x},{\bmath \theta})}{\partial {\bmath \theta}}$. Then $f^{\prime}({\bf x},\widehat{\bmath \theta})={\bf 0}$. Taking derivative of both sides with respect to $x_m$, $\frac{\partial f^\prime}{\partial x_m} \frac{\partial x_m}{\partial x_m}+ \frac{\partial f^\prime}{\partial \widehat{\bmath \theta}} \frac{\partial \widehat{\bmath \theta}}{\partial x_m} ={\bf 0}$. Simplifying this leads to (\ref{theta_der}).}
\beq \label{theta_der}
\frac{\partial \widehat{\bmath \theta}}{\partial x_m} = - \left( f_{\theta \theta} ({\bf x},{\bmath \theta}) \right)^{-1} f_{X_m \theta} ({\bf x},{\bmath \theta}) \arrowvert_{{\bmath \theta}=\widehat{\bmath \theta}}
\eeq
where
\beqn
f_{\theta \theta} ({\bf x},{\bmath \theta}) \buildrel \triangle \over= \frac{\partial^2 f({\bf x},{\bmath \theta})} {\partial {\bmath \theta} \partial{\bmath \theta}^T}   \in \mathbb{R}^{M\times M},\\
f_{X_m \theta} ({\bf x},{\bmath \theta}) \buildrel \triangle \over= \frac{\partial^2 f({\bf x},{\bmath \theta})}{\partial x_m \partial {\bmath \theta}}  \in \mathbb{R}^{M\times 1}.
\eeqn

Consider the mapping from ${\bf x}$ to ${\bf y}$, 
\beq
{\bf y}={\bf T}({\bf x})
\eeq
where ${\bf T}(\cdot)$ is a composite of calibration and model subtraction to get the residual.
Let the joint probability density functions of ${\bf x}$ and ${\bf y}$ be $p_X({\bf x})$ and $p_Y({\bf y})$, respectively.
We can find $p_Y({\bf y})$ by looking at the statistics of the residual, but scientific interest is in finding $p_X({\bf x})$, so we use 
\beq
p_X({\bf x}) = | \mathcal{J} | \ \ p_Y({\bf T}({\bf x}))  
\eeq
where $\mathcal{J} \in \mathbb{R}^{D \times D}$ is the Jacobian of the mapping ${\bf T}(\cdot)$,
\beq \label{Jmapping}
 \mathcal{J} =\left[
\begin{array}{cccc}
\frac{\partial y_1}{\partial x_1} & \frac{\partial y_1}{\partial x_2} & \ldots & \frac{\partial y_1}{\partial x_D}\\
\frac{\partial y_2}{\partial x_1} & \frac{\partial y_2}{\partial x_2} & \ldots & \frac{\partial y_2}{\partial x_D}\\
\vdots & \vdots & \vdots & \vdots\\
\frac{\partial y_D}{\partial x_1} & \frac{\partial y_D}{\partial x_2} & \ldots & \frac{\partial y_D}{\partial x_D}
\end{array} \right].
\eeq
We use (\ref{elem}) to evaluate each column of $\mathcal{J}$. Using (\ref{elem}), we can rewrite (\ref{Jmapping}) as
\beq
 \mathcal{J} ={\bf I}_D +\mathcal{A}
\eeq
where
\beq \label{AA}
 \mathcal{A}\buildrel \triangle\over=\frac{\partial {\bf s}({\bmath \theta})}{\partial {\bmath \theta}^T} \left( f_{\theta \theta} ({\bf x},{\bmath \theta}) \right)^{-1} [f_{X_1 \theta} ({\bf x},{\bmath \theta})\ldots f_{X_D \theta} ({\bf x},{\bmath \theta})] \arrowvert_{{\bmath \theta}=\widehat{\bmath \theta}} 
\eeq

The determinant of $\mathcal{J}$  can be given using eigenvalues of $\mathcal{A}$, i.e. $\lambda_j(\mathcal{A})$ as \cite{Ipsen}
\beq
| \mathcal{J} | = \exp \left(\sum_{j=1}^{D} \log \left(1 + \lambda_j(\mathcal{ A})\right) \right).
\eeq

The evaluation of (\ref{AA}) and moreover its eigenvalues is an expensive task. However, we can use the results of section \ref{sec:perf} to simplify this.
The closed form expressions (\ref{Jderiv}) and (\ref{Rderiv}) can be used to evaluate elements of (\ref{AA}) in closed form. The only requirement is the careful mapping of index $m$ to indices $p^\prime,q^\prime,r$ and vice versa. Moreover, the inversion of $\left( f_{\theta \theta} ({\bf x},{\bmath \theta}) \right)$ in (\ref{AA}) is not explicitly needed because we can use an iterative algorithm to find the eigenvalues of $\mathcal{ A}$ such as by using implicitly restarted Arnoldi methods \cite{ARPACK}.

The extension of this work to study the performance of multi-source calibration is straightforward. We need to partition ${\bmath \theta}$ into partitions corresponding to each direction and we can evaluate (\ref{AA}) in block partitioned form. Furthermore, the results can also be used to study calibration without consensus optimization by setting $\rho=0$ in ((\ref{Jderiv}).

\section{Conclusions}\label{sec:conc}
We have derived closed form relations for the performance analysis of distributed radio interferometric calibration. To study the weak signals buried in the data, preservation of their statistical behavior is essential. Using this work, we are able to study the effect of calibration in possible transformations of input data and if needed, compensating for these effects. We will produce software based on this work to accompany our calibration software as future work.

\begin{appendix}
\noindent{\bf I: Proof of (\ref{gradJ}) and (\ref{gradY})}\\
First note that if $g_1=\| {\bf Y}^H({\bf A}{\bf J} + {\bf B}{\bf Y}+{\bf C})\|$ and  $g_2=\|{\bf A}{\bf J}+{\bf B}{\bf Y}+{\bf C}\|^2$ for some arbitrary constant matrices ${\bf A}$, ${\bf B}$ and ${\bf C}$, then the derivatives with respect to ${\bf J}$ and ${\bf Y}$ are, ${\rm grad}(g_1,{\bf J})=\frac{1}{2}{\bf A}^H{\bf Y}$, ${\rm grad}(g_1,{\bf Y})=\frac{1}{2}({\bf A}{\bf J}+({\bf B}+{\bf B}^H){\bf Y}+{\bf C})$, ${\rm grad}(g_2,{\bf J})={\bf A}^H({\bf A}{\bf J}+{\bf B}{\bf Y}+{\bf C})$ and ${\rm grad}(g_2,{\bf Y})={\bf B}^H({\bf A}{\bf J}+{\bf B}{\bf Y}+{\bf C})$. Using this to find the gradient of (\ref{Lconverge}) we get
\beqn
\lefteqn{{\rm grad}(L_f,{\bf J})}\\\nonumber
&&={\rm grad}(g_{f}({\bf J}_f),{\bf J}_f)+\frac{1}{2}({\bf I}-\rho{\bf B}{\bf P})^H{\bf Y}\\\nonumber
&&+\frac{\rho}{2}({\bf I}-\rho{\bf B}{\bf P})^H\left(({\bf I}-\rho{\bf B}{\bf P}){\bf J}-{\bf B}{\bf P}{\bf Y}-{\bf B}{\bf R}\right)
\eeqn
and
\beqn
\lefteqn{{\rm grad}(L_f,{\bf Y})}\\\nonumber
&&=\frac{1}{2} \left(({\bf I}-\rho{\bf B}{\bf P}){\bf J}-({\bf B}{\bf P}+{\bf P}^H{\bf B}^H){\bf Y}-{\bf B}{\bf R}\right) \\\nonumber
&&-\frac{\rho}{2}({\bf B}{\bf P})^H\left(({\bf I}-\rho{\bf B}{\bf P}){\bf J}-{\bf B}{\bf P}{\bf Y}-{\bf B}{\bf R}\right)
\eeqn
and substitution ${\bf F}={\bf I}-\rho{\bf B}{\bf P}$ leads to (\ref{gradJ}) and (\ref{gradY}).

\noindent{\bf II: Proof of derivative (\ref{Jderiv})}\\
Taking the differential of (\ref{gradY}) at the solution (first using ${\rm grad}(L_f,{\bf Y})={\bf 0}$)
\beq
d {\bf Y}=\rho ({\bf I}-{\bf F}^H{\bf F})^{-1}{\bf F}^H{\bf F} d {\bf J}
\eeq
and substituting this to the differential of  (\ref{gradJ}) at the solution (using ${\rm grad}(L_f,{\bf J})={\bf 0}$)
\beqn
d {\rm grad}(g_{f}({\bf J}_f),{\bf J}_f) &&+ \frac{\rho}{2} {\bf F}^H{\bf F}\left({\bf I} \right.\\\nonumber
&&+\left. \left({\bf I}-{\bf F}^H{\bf F}\right)^{-1} \right) d {\bf J}={\bf 0}
\eeqn
We use the chain rule to expand $d {\rm grad}(g_{f}({\bf J}_f),{\bf J}_f)$ as
\beqn
d \mathrm{vec}\left({\rm grad}(g_{f}({\bf J}_f),{\bf J}_f)\right) = \mathcal{D}_{\bf J}{\rm grad}(g_{f}({\bf J}_f)) \mathrm{vec}\left(d{\bf J})\right)\\\nonumber
 +\frac{\partial}{\partial x_{p^\prime q^\prime r}} \mathrm{vec}\left({\rm grad}(g_{f}({\bf J}_f),{\bf J}_f) \right)
\eeqn
where $\mathcal{D}_{\bf J}{\rm grad}(g_{f}({\bf J}_f))$ is found by using definition 4 of \cite{Mdiff}. In other words, if ${\bf G}({\bf J},{\bf J}^{\star})$ is a matrix function of ${\bf J}$, the derivatives satisfy $d \mathrm{vec}\left({\bf G}\right) = \left(\mathcal{D}_{\bf J} {\bf G}\right) d\mathrm{vec}\left({\bf J}\right) + \left(\mathcal{D}_{{\bf J}^\star} {\bf G}\right) d\mathrm{vec}\left({\bf J}^\star\right)$ and what we need is $\left(\mathcal{D}_{\bf J} {\bf G}\right)$.
\end{appendix}
\bibliographystyle{IEEEtran}
\bibliography{references}

\begin{thebibliography}{10}
\providecommand{\url}[1]{#1}
\csname url@samestyle\endcsname
\providecommand{\newblock}{\relax}
\providecommand{\bibinfo}[2]{#2}
\providecommand{\BIBentrySTDinterwordspacing}{\spaceskip=0pt\relax}
\providecommand{\BIBentryALTinterwordstretchfactor}{4}
\providecommand{\BIBentryALTinterwordspacing}{\spaceskip=\fontdimen2\font plus
\BIBentryALTinterwordstretchfactor\fontdimen3\font minus
  \fontdimen4\font\relax}
\providecommand{\BIBforeignlanguage}[2]{{%
\expandafter\ifx\csname l@#1\endcsname\relax
\typeout{** WARNING: IEEEtran.bst: No hyphenation pattern has been}%
\typeout{** loaded for the language `#1'. Using the pattern for}%
\typeout{** the default language instead.}%
\else
\language=\csname l@#1\endcsname
\fi
#2}}
\providecommand{\BIBdecl}{\relax}
\BIBdecl

\bibitem{SZ2013}
S.~{Zaroubi}, ``{The Epoch of Reionization},'' in \emph{The First Galaxies},
  ser. Astrophysics and Space Science Library, T.~{Wiklind}, B.~{Mobasher}, and
  V.~{Bromm}, Eds., vol. 396, 2013, p.~45.

\bibitem{Barry2016}
N.~{Barry}, B.~{Hazelton}, I.~{Sullivan}, M.~F. {Morales}, and J.~C. {Pober},
  ``{Calibration requirements for detecting the 21 cm epoch of reionization
  power spectrum and implications for the SKA},'' \emph{\mnras}, vol. 461, pp.
  3135--3144, Sep. 2016.

\bibitem{boyd2011}
S.~Boyd, N.~Parikh, E.~Chu, B.~Peleato, and J.~Eckstein, ``Distributed
  optimization and statistical learning via the alternating direction method of
  multipliers,'' \emph{Foundations and Trends{\textregistered} in Machine
  Learning}, vol.~3, no.~1, pp. 1--122, 2011.

\bibitem{DCAL}
S.~{Yatawatta}, ``Distributed radio interferometric calibration,''
  \emph{\mnras}, vol. 449, no.~4, pp. 4506--4514, 2015.

\bibitem{EUSIPCO2016}
S.~Yatawatta, ``Fine tuning consensus optimization for distributed radio
  interferometric calibration,'' in \emph{2016 24th European Signal Processing
  Conference (EUSIPCO)}, Aug 2016, pp. 265--269.

\bibitem{Brossard2018}
\BIBentryALTinterwordspacing
M.~Brossard, M.~N.~E. Korso, M.~Pesavento, R.~Boyer, P.~Larzabal, and S.~J.
  Wijnholds, ``Parallel multi-wavelength calibration algorithm for radio
  astronomical arrays,'' \emph{Signal Processing}, vol. 145, pp. 258 -- 271,
  2018. [Online]. Available:
  \url{http://www.sciencedirect.com/science/article/pii/S0165168417304279}
\BIBentrySTDinterwordspacing

\bibitem{CAMSAP2017}
S.~Yatawatta, F.~Diblen, and H.~Spreeuw, ``Adaptive {ADMM} in distributed radio
  interferometric calibration,'' in \emph{2017 IEEE 7th International Workshop
  on Computational Advances in Multi-Sensor Adaptive Processing (CAMSAP)}, Dec
  2017, pp. 1--5.

\bibitem{DMUX}
S.~{Yatawatta}, F.~{Diblen}, H.~{Spreeuw}, and L.~V.~E. {Koopmans}, ``{Data
  multiplexing in radio interferometric calibration},'' \emph{\mnras}, vol.
  475, pp. 708--715, Mar. 2018.

\bibitem{Onose}
A.~{Onose}, R.~E. {Carrillo}, A.~{Repetti}, J.~D. {McEwen}, J.-P. {Thiran},
  J.-C. {Pesquet}, and Y.~{Wiaux}, ``{Scalable splitting algorithms for
  big-data interferometric imaging in the SKA era},'' \emph{ArXiv e-prints},
  Jan. 2016.

\bibitem{Meil2016}
C.~Meillier, P.~Bianchi, and W.~Hachem, ``Two distributed algorithms for the
  deconvolution of large radio-interferometric multispectral images,'' in
  \emph{2016 24th European Signal Processing Conference (EUSIPCO)}, Aug 2016,
  pp. 728--732.

\bibitem{Degu2016}
J.~Deguignet, A.~Ferrari, D.~Mary, and C.~Ferrari, ``Distributed
  multi-frequency image reconstruction for radio-interferometry,'' in
  \emph{2016 24th European Signal Processing Conference (EUSIPCO)}, Aug 2016,
  pp. 1483--1487.

\bibitem{Onose2017}
A.~{Onose}, A.~{Dabbech}, and Y.~{Wiaux}, ``{An accelerated splitting algorithm
  for radio-interferometric imaging: when natural and uniform weighting
  meet},'' \emph{ArXiv e-prints}, Jan. 2017.

\bibitem{Zmu}
J.~Zmuidzinas, ``Cram\'{e}r--{R}ao sensitivity limits for astronomical
  instruments: implications for interferometer design,'' \emph{Journal of the
  Optical Society of America A}, vol.~20, no.~2, pp. 218--233, Feb 2003.

\bibitem{Jeffs06}
S.~{van der Tol}, B.~{Jeffs}, and A.~{van der Veen}, ``Self calibration for the
  {LOFAR} radio astronomical array,'' \emph{IEEE Trans. Sig. Proc.}, vol.~55,
  no.~9, pp. 4497--4510, Sep. 2007.

\bibitem{Wijn}
S.~{Wijnholds} and A.~{van der Veen}, ``Multisource self-calibration for sensor
  arrays,'' \emph{IEEE Trans. Sig. Proc.}, vol.~57, no.~9, pp. 3512--3532, May
  2009.

\bibitem{Kazemi12}
S.~Kazemi, S.~Yatawatta, and S.~Zaroubi, ``Performance analysis of clustered
  radio interferometric calibration,'' in \emph{Acoustics, Speech and Signal
  Processing (ICASSP), 2012 IEEE International Conference on}, March 2012, pp.
  2533--2536.

\bibitem{cook1982residuals}
\BIBentryALTinterwordspacing
R.~Cook and S.~Weisberg, \emph{Residuals and Influence in Regression}, ser.
  Monographs on statistics and applied probability.\hskip 1em plus 0.5em minus
  0.4em\relax Chapman \& Hall, 1982. [Online]. Available:
  \url{http://books.google.nl/books?id=MVSqAAAAIAAJ}
\BIBentrySTDinterwordspacing

\bibitem{Laurent92}
R.~T. {St. Laurent} and R.~D. {Cook}, ``{Leverage, and superleverage in
  nonlinear regression},'' \emph{Journal of the American Statistical
  Association}, vol.~87, no. 420, pp. 985--990, 1992.

\bibitem{Laurent93}
------, ``{Leverage, local influence and curvature in nonlinear regression},''
  \emph{Biometrika}, vol.~80, no.~1, pp. 99--106, 1993.

\bibitem{SS9}
S.~Yatawatta, ``Jacobian leverage as a diagnostic in radio interferometric
  calibration,'' in \emph{Radio Science Conference (URSI AT-RASC), 2015 1st
  URSI Atlantic}, May 2015, pp. 1--1.

\bibitem{Patil2016}
A.~H. {Patil}, S.~{Yatawatta}, S.~{Zaroubi}, L.~V.~E. {Koopmans}, A.~G. {de
  Bruyn}, V.~{Jeli{\'c}}, B.~{Ciardi}, I.~T. {Iliev}, M.~{Mevius}, V.~N.
  {Pandey}, and B.~K. {Gehlot}, ``{Systematic biases in low-frequency radio
  interferometric data due to calibration: the LOFAR-EoR case},''
  \emph{\mnras}, vol. 463, pp. 4317--4330, Dec. 2016.

\bibitem{Samuel}
K.~G.~G. Samuel and M.~F. Tappen, ``Learning optimized {MAP} estimates in
  continuously-valued {MRF} models,'' in \emph{2009 IEEE Conference on Computer
  Vision and Pattern Recognition}, June 2009, pp. 477--484.

\bibitem{Gould2016}
S.~{Gould}, B.~{Fernando}, A.~{Cherian}, P.~{Anderson}, R.~{Santa Cruz}, and
  E.~{Guo}, ``{On Differentiating Parameterized Argmin and Argmax Problems with
  Application to Bi-level Optimization},'' \emph{ArXiv e-prints}, Jul. 2016.

\bibitem{Mdiff}
A.~{Hjorungnes} and D.~{Gesbert}, ``Complex valued matrix differentiation:
  Techniques and key results,'' \emph{IEEE Trans. on Sig. Proc.}, vol. 55, no.
  6, pp. 2740--2746, Jun. 2007.

\bibitem{HBS}
J.~P. Hamaker, J.~D. Bregman, and R.~J. Sault, ``{Understanding radio
  polarimetry, paper I},'' \emph{Astronomy and Astrophysics Supp.}, vol. 117,
  no. 137, pp. 96--109, 1996.

\bibitem{TMS}
A.~{Thompson}, J.~{Moran}, and G.~{Swenson}, \emph{{Interferometry and
  synthesis in radio astronomy (3rd ed.)}}.\hskip 1em plus 0.5em minus
  0.4em\relax New York: Wiley Interscience, 2001.

\bibitem{Kaz3}
S.~{Kazemi} and S.~{Yatawatta}, ``{Robust radio interferometric calibration
  using the t-distribution},'' \emph{\mnras}, vol. 435, pp. 597--605, Oct.
  2013.

\bibitem{SIRP}
V.~{Ollier}, M.~N.~E. {Korso}, R.~{Boyer}, P.~{Larzabal}, and M.~{Pesavento},
  ``{Robust Calibration of Radio Interferometers in Non-Gaussian
  Environment},'' \emph{IEEE Transactions on Signal Processing}, vol.~65, pp.
  5649--5660, Nov. 2017.

\bibitem{grobler2014}
T.~Grobler, C.~Nunhokee, O.~Smirnov, A.~Van~Zyl, and A.~De~Bruyn, ``Calibration
  artefacts in radio interferometry--i. ghost sources in {W}esterbork synthesis
  radio telescope data,'' \emph{Monthly Notices of the Royal Astronomical
  Society}, vol. 439, no.~4, pp. 4030--4047, 2014.

\bibitem{Ipsen}
I.~C.~F. {Ipsen} and D.~J. {Lee}, ``{Determinant Approximations},'' \emph{ArXiv
  e-prints}, May 2011.

\bibitem{ARPACK}
\BIBentryALTinterwordspacing
R.~Lehoucq, D.~Sorensen, and C.~Yang, \emph{ARPACK Users' Guide}.\hskip 1em
  plus 0.5em minus 0.4em\relax Society for Industrial and Applied Mathematics,
  1998. [Online]. Available:
  \url{https://epubs.siam.org/doi/abs/10.1137/1.9780898719628}
\BIBentrySTDinterwordspacing

\end{thebibliography}

\end{document}